\documentclass[]{tPHM2e}

\begin{document}
\doi{10.1080/14786435.20xx.xxxxxx}
\issn{1478-6443}
\issnp{1478-6435}
\jvol{00} \jnum{00} \jyear{2011} 

\markboth{Sheng Ran, Sergey L. Bud'ko and  Paul C. Canfield}{Philosophical Magazine}

\articletype{}

\title{Effects of substitution on low temperature physical properties of LuFe$_{2}$Ge$_{2}$}

\author{Sheng Ran, Sergey L. Bud'ko and  Paul C. Canfield$^{\ast}$\thanks{$^\ast$Corresponding author. Email: canfield@ameslab.gov
\vspace{6pt} }  \\\vspace{6pt} 
{\em{Ames Laboratory, US DOE, and Department of Physics and Astronomy, Iowa State University, Ames, IA 50011, USA}}; \\\vspace{6pt}\received{Received 04 July 2011} }

\maketitle

\begin{abstract}
The low temperature, magnetic phase transition in LuFe$_{2}$Ge$_{2}$ is thought to be associated with itinerant magnetism.  The effects of Y and Sc substitutions on the Lu site, as well as Ru and Co substitutions on the Fe site, on the low temperature magnetic phase transition of LuFe$_{2}$Ge$_{2}$ compound have been studied in single crystals via microscopic, thermodynamic and transport measurements.  On one hand, Co substitution suppresses the transition below our base temperature of 2~K even at our lowest substitution level.  On the other hand, Sc substitution enhances the transition temperature, and Y or Ru substitution suppresses the transition to lower temperature.  Phase diagrams for Y, Sc and Ru substitutions have been constructed and the possibility of a unifying, composite diagram is discussed.  

\bigskip

\end{abstract}

\section{Introduction}

In the recently discovered iron-based superconductors substitutions to the parent compounds that add electrons or holes, as well as isoelectronic substitutions and pressure, suppress the structural and magnetic phase transitions and can ultimately reveal superconductivity with a relatively high transition temperature \cite{JPSJFeAs,Hosono09,Chu09,Prozorov10,Canfield10}. The proximity of the suppressed magnetic / structural phase transitions to the maximum {\itshape T}$_{c}$ values, as well as more direct evidence, suggest the importance of magnetism, most likely itinerant magnetism, to the superconducting state \cite{Canfield10}.  Among the Fe-pnictide-based superconductors, one of the most extensively studied families, AEFe$_{2}$As$_{2}$, (AE = Ca, Sr, Ba) forms in ThCr$_{2}$Si$_{2}$ crystal structure with space group I4/mmm. 

ThCr$_{2}$Si$_{2}$-type compounds are some of the most common ternary intermetallic phases; for example, the RT$_{2}$X$_{2}$ (R = Y, La$-$Lu; T = Mn$-$Cu, Ru, Rh, etc. and X = Si, Ge) series have been intensively studied for several decades due to the wide range of exciting physical properties displayed by their members \cite{Szytula89}.  Out of all of the transition metals, T, only Mn carries a local moment with the RMn$_{2}$Ge$_{2}$ and RMn$_{2}$Si$_{2}$ compounds ordering magnetically at greatly enhanced temperatures relative to the other RT$_{2}$X$_{2}$ series.  It is worth noting, though, that LuFe$_{2}$Ge$_{2}$ manifests anomalies in susceptibility, resistivity and specific heat at 9~K that have been associated with itinerate magnetic order \cite{Avila04}. Although the precise nature of the magnetic ordering is still unclear, analysis of susceptibility, resistivity and specific heat data lead to the prediction of a SDW state, most likely with an ordering wave vector along a [00l] direction, a result subsequently found by neutron scattering measurements \cite{Fujiwara07}. Two alternate hypotheses are (i) that LuFe$_{2}$Ge$_{2}$ (and YFe$_{2}$Ge$_{2}$) are close to the Stoner limit and easily forced into a magnetically ordered state, or (ii) that the Fe is moment bearing with a large paramagnetic effective moment \cite{Ferstl06}, but both of these hypotheses are inconsistent with the relatively low ordering temperatures of the other RFe$_{2}$Ge$_{2}$ members which are closer to those of RNi$_{2}$Ge$_{2}$ \cite{Budko99} rather than RMn$_{2}$Ge$_{2}$ \cite{Szytula89} or the Stoner enhanced RFe$_{2}$Ge$_{20}$ compounds \cite{Jia07,JiaPRB07,Jia08}.

In order to better characterize this phase transition, in this work we report the effects of Y and Sc substitutions on the Lu site, as well as Ru and Co substitutions on the Fe site, on the low temperature properties of the parent compound and present the phase diagram for Y, Sc and Ru substitutions. The isoelectronic substitution: Y, Sc, and Ru, can be considered as chemical pressure or strain causing primarily steric effects, whereas Co substitution, can be considered (in a rigid band model) as adding electrons and thereby causing changes in the band filling as well. The goal of this work was to see how the antiferromagnetic ordering evolves with different substitutions.

\section{Experimental methods}

Single crystals of pure and substituted LuFe$_{2}$Ge$_{2}$ were grown out of Sn flux \cite{Avila04,Fisk89,Canfield92,Canfield01,Canfieldbook}. A typical procedure involved adding, to a 2~ml alumina crucible, about 5~g of Sn, and roughly 6~at$\%$ of LuFe$_{2}$Ge$_{2}$ in elemental form. For Sc and Y substitutions, elements were mixed together according to the ratio Lu:Sc/Y:Fe:Ge:Sn = 1-{\itshape x}:{\itshape x}:2.4:2:90, where {\itshape x} is the nominal concentration of the substitutions. For Co and Ru substitutions, the ratio was Lu:Fe:Co/Ru:Ge:Sn = 1:2.4(1-{\itshape x}):2.4{\itshape x}:2:90. The 20$\%$ excess of Fe was used to suppress the growth of a second phase, LuFe$_{6}$Ge$_{6}$, although it was not crucial and did not change the actual measured properties of the LuFe$_{2}$Ge$_{2}$ crystals. (As indicated by $M(H)$ measurements, no additional ferromagnetic impurities were induced due to the extra Fe in the starting material.)  The crucible, with starting elements, was sealed in a fused silica ampoule under partial argon atmosphere and then placed in a box furnace and heated up to 1190~$^\circ$C over 6~h and held at 1190~$^\circ$C for 2~h. The crystals grew while the temperature was reduced to 550~$^\circ$C over 96~h, after which the ampoule was quickly removed from the furnace and the molten, Sn-rich solution was decanted by use of a centrifuge \cite{Fisk89,Canfield92,Canfield01,Canfieldbook}. A concentrated HCl etch was used to remove residual Sn from the crystal surfaces.

Elemental analysis was performed on each of these batches using wavelength-dispersive x-ray spectroscopy (WDS) in the electron probe microanalyzer of a JEOL JXA-8200 electron microprobe. To get flat surfaces for WDS analysis, all samples were carefully polished by embedding the samples in epoxy in 0.5 cm diameter carbon ring forms and polishing the surfaces with sand papers and alumina abrasives until a final surface polish with $\backsim$ 1~$\mu$m roughness was achieved. WDS measurements were done at 12 locations of samples from each batch. 

Powder x-ray diffraction measurements, with a Si standard, were performed at room temperature using a Rigaku Miniflex diffractometer with Cu {\itshape K}$\alpha$ radiation. Diffraction patterns were taken on ground single crystals from each batch. The unit cell parameters were refined by Rietica software. We analyzed more than one set of data for some representative substitutions and there is scatter in lattice parameters determined from different sets of data. Therefore the error bars were taken as half of the maximum spread. It turns out that the relative error bar in {\itshape c}-lattice parameter is roughly twice of that in {\itshape a}-lattice parameter.

\begin{figure}
\begin{center}
\resizebox*{10cm}{!}{\includegraphics{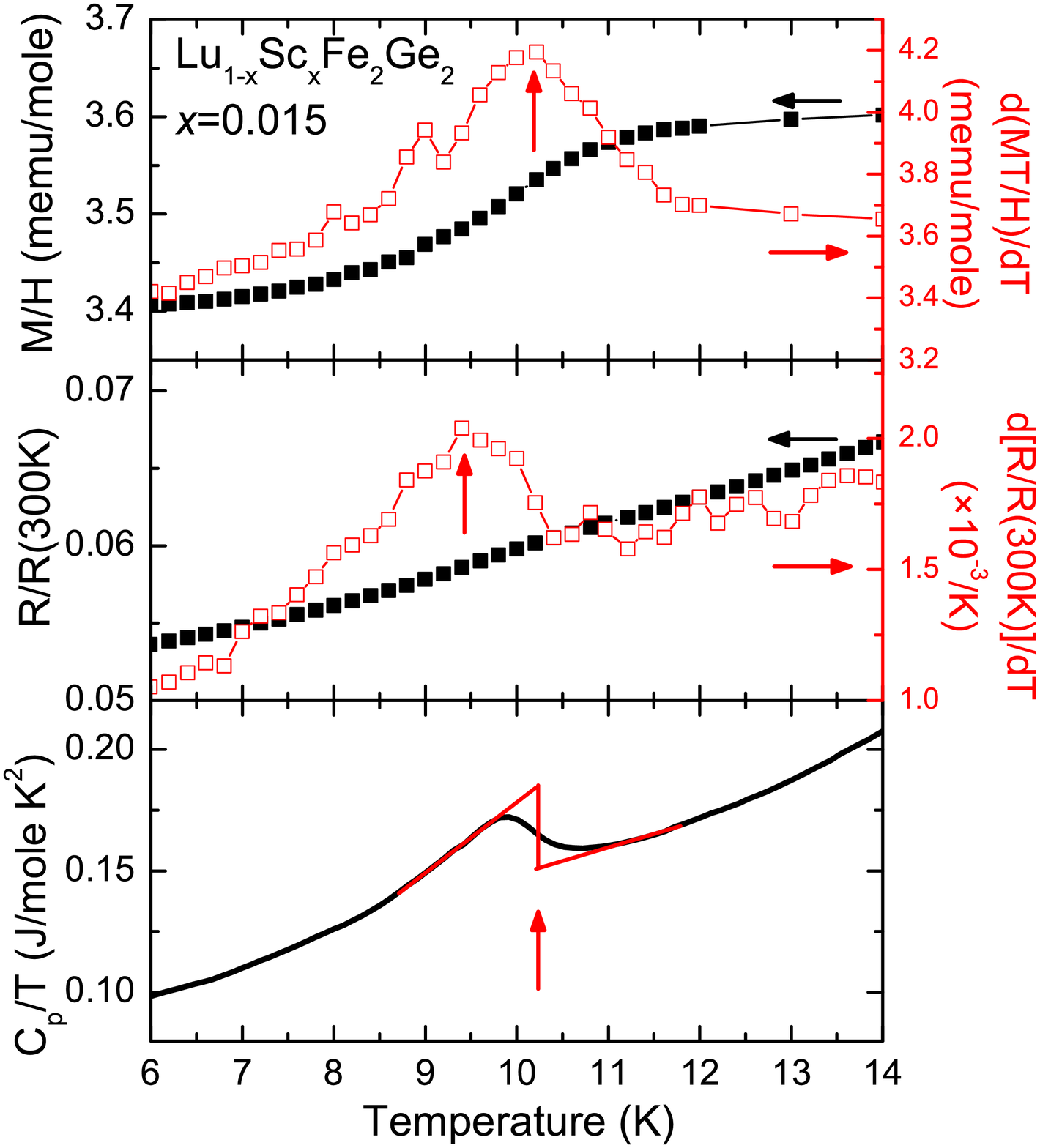}}%
\caption{ Criteria used to determine values for transition temperatures. Upper panel: $M/H$ and $d(MT/H)/dT$ with a field of 1T applied parallel to the crystallographic ab plane. Middle panel: $R/R(300K)$ and $d[R/R(300K)]/dT$. Bottom panel: the approximate isoentropic construction of specific heat.}%
\label{criteria}
\end{center}
\end{figure}

Temperature dependent magnetization measurements were made in Quantum Design MPMS systems. The in plane temperature dependent electrical AC ({\itshape f} = 16~Hz, {\itshape I} = 1~mA) resistivity measurements were performed in Quantum Design MPMS systems operated in external device control (EDC) mode, in conjunction with Linear Research LR700 four-probe AC resistance bridges. The electrical contacts were placed on the samples in standard 4-probe geometry, using Pt wires attached to a sample surface with Epotek H20E silver epoxy. The room temperature resistivity of LuFe$_{2}$Ge$_{2}$ is $\backsim$ 250~$\mu\Omega$-cm and does not vary significantly with these small substitutions. For clarity of comparison, resistivity data will be shown as $R(T)/R(300K)$. Temperature dependent heat capacity for representative samples was measured in a Quantum Design PPMS system using the relaxation technique in zero applied magnetic field.  Transition temperatures were inferred from the peak of $d(MT/H)/dT$ and $d[R/R(300K)]/dT$ \cite{Fisher62,Fisher68} ($M,H,R,T$ stand for magnetization, applied field, resistance and temperature, respectively) as well as the approximate isoentropic construction in heat capacity. The criteria used to determine transition temperatures are shown in Figure \ref{criteria} for a representative Sc substitution level of {\itshape x} = 0.015. 

\section{Results}

A summary of the WDS measurement data is presented in Table 1. The table shows the nominal concentration, the measured average {\itshape x} values, and two times the standard deviation of the {\itshape x} values measured. For each substitution, data points of nominal versus actual concentration can be fitted very well with straight lines, with slopes of 1.08 $\pm$ 0.03, 0.23 $\pm$ 0.005, 0.70 $\pm$ 0.01 and 1.75 $\pm$ 0.07 for Y, Sc, Ru and Co substitution, respectively. It can be seen that the difference between nominal and WDS concentration is very different for different substitutions. The nearly linear dependence indicates a close correlation between the measured substitution concentration and the nominal concentration. The error bars are taken as twice of the standard deviation determined from the measurements. The compositional spread over the sample surfaces for each concentration is no more than 0.015, demonstrating relative homogeneity of the substitution studied here. (For the lowest Sc, Ru, and Co substitution levels the 2$\sigma$ values were 0.002 or less.) In the following, the average experimentally determined {\itshape x} values, {\itshape x}$_{WDS}$, will be used to identify all the compounds rather than the nominal concentration, {\itshape x}$_{nominal}$. 

\begin{table}
  \tbl{WDS data for all four series. {\itshape x}$_{nominal}$ is the nominal concentration of the substitutions. {\itshape x}$_{WDS}$ is the average {\itshape x} values measured at 12 locations of samples in each batch. 2$\sigma$ is two times the standard deviation of the 12 values measured.}
{\begin{tabular}{@{}lcccccc}
\toprule
    \multicolumn{7}{c}{Lu$_{1-x}$Y$_{x}$Fe$_{2}$Ge$_{2}$} \\ 
\colrule
   {\itshape x}$_{nominal}$ & 0.05	& 0.10	& 0.13	&0.16	&0.19	&0.50 \\
   {\itshape x}$_{WDS}$ &0.043	&0.094	&0.125	&0.148	&0.19	&0.56 \\
   2$\sigma$	&0.006	&0.007	&0.005	&0.005	&0.01	&0.016 \\
\toprule
\multicolumn{7}{c}{Lu$_{1-x}$Sc$_{x}$Fe$_{2}$Ge$_{2}$ } \\ 
\colrule
   {\itshape x}$_{nominal}$ &0.03	&0.06	&0.10	&0.20\\
   {\itshape x}$_{WDS}$ &0.008	&0.015	&0.024	&0.045 \\
   2$\sigma$	&0.002	&0.002	&0.006	&0.006 \\
\toprule
\multicolumn{7}{c}{Lu(Fe$_{1-x}$Ru$_{x}$)$_{2}$Ge$_{2}$} \\ 
\colrule
   {\itshape x}$_{nominal}$ &0.01	&0.02	&0.05\\
   {\itshape x}$_{WDS}$ &0.008	&0.014	&0.035 \\
   2$\sigma$	&0.001	&0.002	&0.001 \\
\toprule
\multicolumn{7}{c}{Lu(Fe$_{1-x}$Co$_{x}$)$_{2}$Ge$_{2}$ } \\ 
\colrule
   {\itshape x}$_{nominal}$ &0.01	&0.02	&0.025	&0.05	&0.10	&0.20\\
   {\itshape x}$_{WDS}$ &0.018	&0.034	&0.056	&0.11	&0.156	&0.33\\
   2$\sigma$	&0.001	&0.002	&0.005	&0.01	&0.004	&0.01\\
\botrule
  \end{tabular}}

\end{table}

\begin{figure}
\begin{center}
\resizebox*{14cm}{!}{\includegraphics{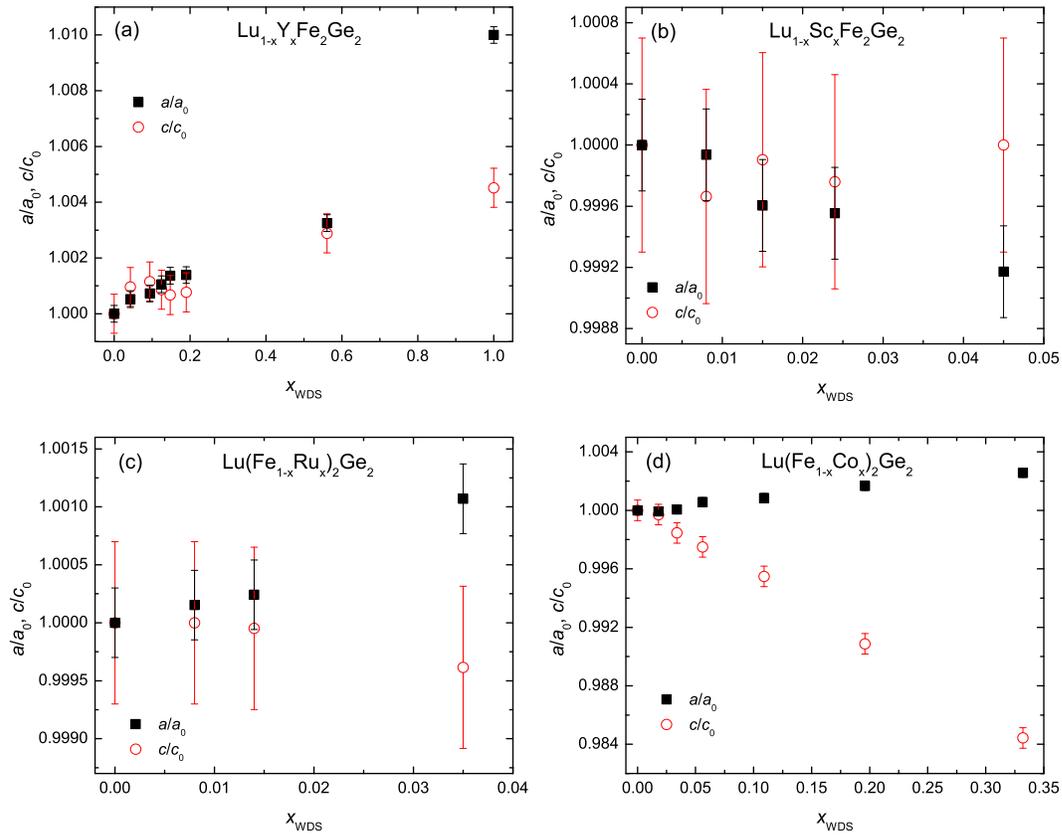}}%
\caption{ Room temperature {\itshape a} and {\itshape c} lattice parameters of (a) the Lu$_{1-x}$Y$_{x}$Fe$_{2}$Ge$_{2}$ series, (b) the Lu$_{1-x}$Sc$_{x}$Fe$_{2}$Ge$_{2}$ series, (c) the Lu(Fe$_{1-x}$Ru$_{x}$)$_{2}$Ge$_{2}$ series and (d) the Lu(Fe$_{1-x}$Co$_{x}$)$_{2}$Ge$_{2}$series, normalized to {\itshape a}$_{0}$=3.9253~\AA\ and {\itshape c}$_{0}$=10.405~\AA\  of pure LuFe$_{2}$Ge$_{2}$  as a function of measured substitutionts concentration, {\itshape x}$_{WDS}$.}%
\label{lattice}
\end{center}
\end{figure}

Figure \ref{lattice} shows the lattice parameters {\itshape a} and {\itshape c} for different substitutions as a function of {\itshape x}$_{WDS}$. For Y substitution, lattice parameter {\itshape a} increases in a roughly linear manner as {\itshape x}$_{WDS}$. Lattice parameter {\itshape c} also increases but with larger scatter in the data for low substitution levels. For Sc substitution, lattice parameter {\itshape a} decreases with {\itshape x}$_{WDS}$ whereas lattice parameter {\itshape c} remains constant within the error bars. For Co substitution, the lattice parameter {\itshape c} decreases with {\itshape x}$_{WDS}$, whereas the lattice parameter {\itshape a} remains almost unchanged at low substitution level then increases slightly at high substitution levels. For Ru substitution, the lattice parameter {\itshape a} increases whereas the lattice parameter {\itshape c} decreases. For all Y, Sc and Ru substitutions, it appears that the lattice parameter {\itshape c} is less sensitive to the substitution than lattice parameter {\itshape a}. In addition, the error bar in lattice parameter {\itshape c} is roughly twice of that in lattice parameter {\itshape a}, making it difficult to determine the changes in the lattice parameter {\itshape c} precisely.

\begin{figure}
\begin{center}
\resizebox*{14cm}{!}{\includegraphics{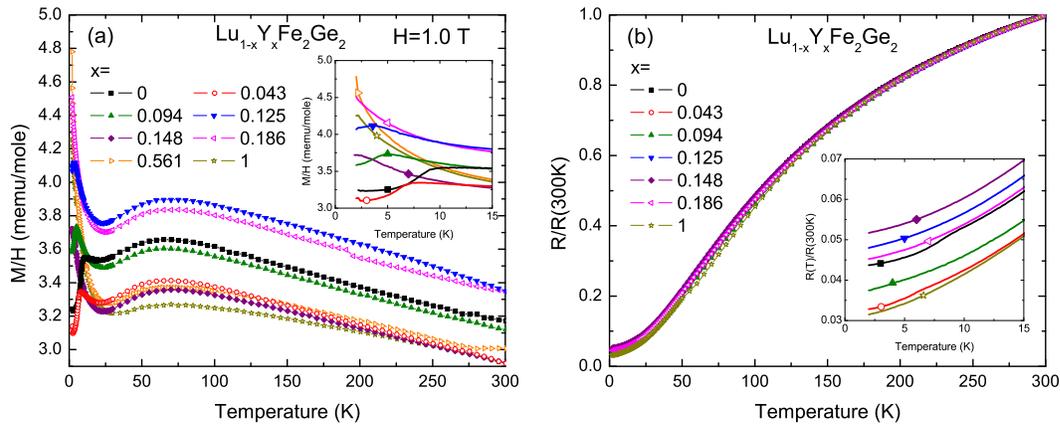}}%
\caption{ Temperature dependent (a) magnetization divided by applied field with a field of 1T applied parallel to the crystallographic ab plane and (b) normalized electrical resistivity of the Lu$_{1-x}$Y$_{x}$Fe$_{2}$Ge$_{2}$ series. Insets show data at low temperature. Transition temperatures are determined using the criteria described in the text.}%
\label{YT}
\end{center}
\end{figure}

Figure \ref{YT}a shows the temperature dependent magnetization data for H $\parallel$ ab of the Lu$_{1-x}$Y$_{x}$Fe$_{2}$Ge$_{2}$ series which was measured in the field of 1~T. The parent compound, LuFe$_{2}$Ge$_{2}$, shows a weak temperature dependence that is consistent with a somewhat enhanced Pauli paramagnetic behavior \cite{Avila04}, but that has also been fit to a Curie Weiss behavior, albeit with an unrealistically high paramagnetic $\theta$ of $\backsim$ 800~K \cite{Ferstl06}.  Upon cooling to low temperatures there is a clear local maximum followed by a sharp drop; analysis of $d(MT/H)/dT$ gives a transition temperature of 8.2~K, a value that is similar to, but somewhat lower than, the previous report of 9~K \cite{Avila04}. By substituting Y onto the Lu site, this transition is suppressed to lower temperatures, ultimately dropping below 2~K for {\itshape x} $\textgreater$ 0.148. The signature of the transition evolves gradually with the substitution level.  As the transition is suppressed a clear, low temperature minimum in $M(T)/H$ is revealed, followed at lowest temperatures by a sharp upturn.  By {\itshape x} = 0.19 the form of $M(T)/H$ is essentially identical to that of pure YFe$_{2}$Ge$_{2}$.  It is worth noting that this lowest temperature tail does not seem to be extrinsic since it essentially disappears below the tunable magnetic transition.  

The effect of Y substitution on the low temperature properties of Lu$_{1-x}$Y$_{x}$Fe$_{2}$Ge$_{2}$ compounds can also be seen in the electrical transport data which is shown in Figure \ref{YT}b. The slope of $R(T)$ for parent compound changes slightly at around 8~K, which corresponds to the anomaly seen in the magnetization data. Upon Y substitution, the resistive feature remains weak and becomes difficult to resolve for {\itshape x} = 0.125, even though the magnetization data show a clear anomaly centered at 3.4~K. For {\itshape x} $\geq$ 0.148, the signatures in both magnetization and resistivity are suppressed completely.

\begin{figure}
\begin{center}
\resizebox*{14cm}{!}{\includegraphics{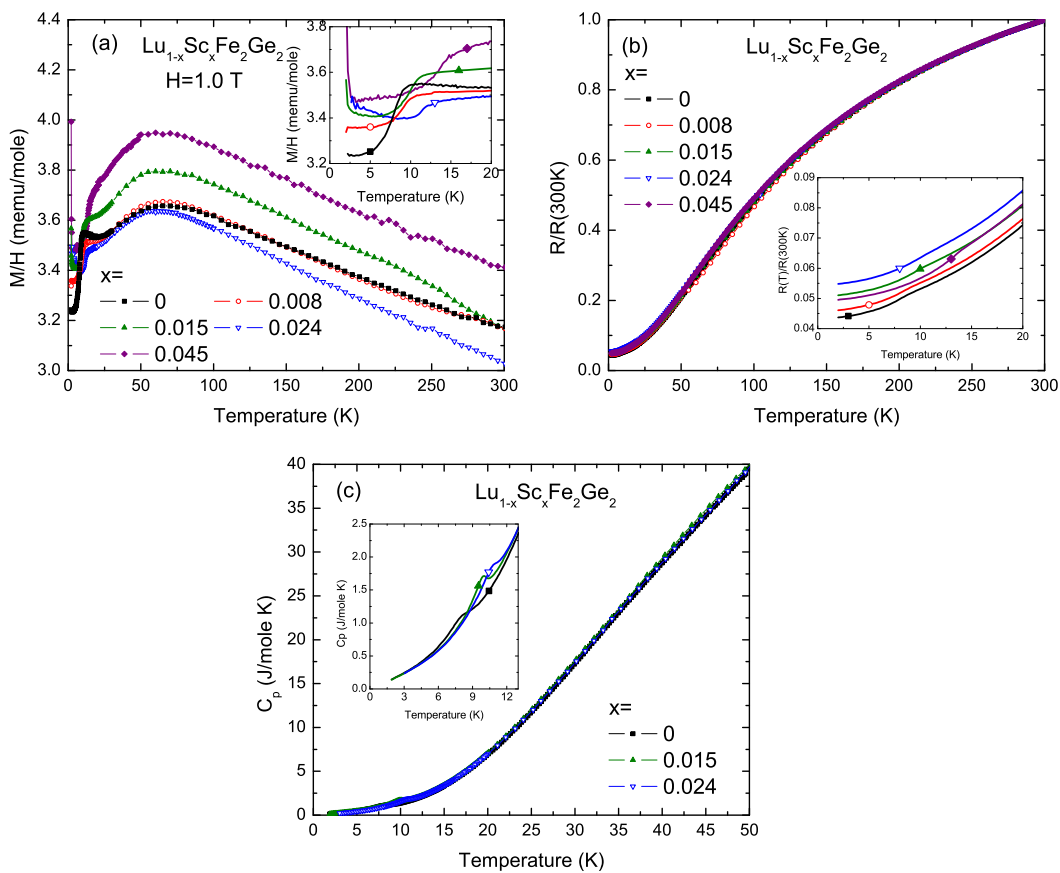}}%
\caption{ Temperature dependent (a) magnetization divided by applied field with a field of 1T applied parallel to the crystallographic ab plane, (b) normalized electrical resistivity and (c) specific heat of the Lu$_{1-x}$Sc$_{x}$Fe$_{2}$Ge$_{2}$ series. Insets show data at low temperature. Transition temperatures are determined using the criteria described in the text.}%
\label{ScT}
\end{center}
\end{figure}

The effect of Sc substitution on the Lu site on the low temperature properties is markedly different from that of Y substitution as manifested by the temperature dependent magnetization and resistivity data as shown in Figures \ref{ScT}a and \ref{ScT}b. As the Sc substitution level increases, the signatures in both magnetization and resistivity are pushed up to higher temperatures instead of being suppressed. Whereas the form of the resistive signature remains essentially unchanged (a weak decrease in resistance similar to a minor reduction in scattering), the magnetic signature evolves in a way different from that of Y substituted compound.  The weak local minimum in the susceptibility, seen for temperatures just above the magnetic transition disappears as the magnetic ordering temperature increases; ultimately, for the highest Sc substitution level, {\itshape x} = 0.045, the sharp drop in susceptibility associated with the magnetic transition occurs abruptly without any hint of a local minima in $M(T)/H$. The enhancement of the transition temperature is further confirmed by specific heat measurement on samples with selected substitution levels (Figure \ref{ScT}c). It can be seen that the corresponding anomaly in specific heat is small but well resolved. With increasing the Sc substitution level, the anomaly shifts to higher temperature. It would be interesting to see the evolution of the transition temperature as well as the signatures of the transition at higher substitution level. Unfortunately, as the substitution level increases, a second phase with different crystal morphology begins to grow and becomes more and more pervasive. Already the nominal {\itshape x} = 0.20 growth, our highest substitution in this work, yields mostly this second phase and only a small amount of clean 122 phase that had to be carefully separated. 

The enhancement of transition temperature by Sc substitution as well as the suppression of transition by Y substitution is consistent with the result of an existing pressure study of LuFe$_{2}$Ge$_{2}$ \cite{Fujiwara07} which shows that on applying pressure, the transition temperature monotonically increases.  With Sc substitution for Lu, both lattice parameter {\itshape a} and {\itshape c} decrease, indicating that Sc substitution serves as a chemical pressure. On the other hand, Y substitution leads to increases in both {\itshape a} and {\itshape c} lattice parameters, making it similar to negative pressure.

\begin{figure}
\begin{center}
\resizebox*{14cm}{!}{\includegraphics{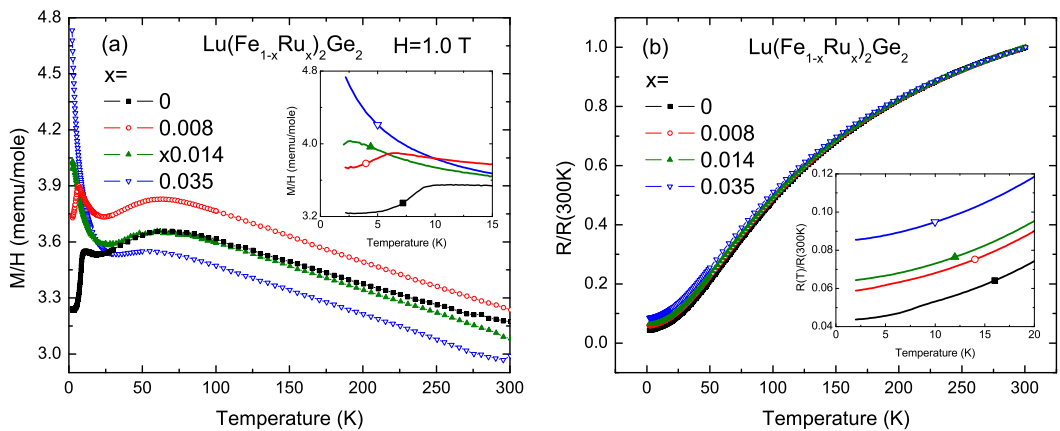}}%
\caption{ Temperature dependent (a) magnetization divided by applied field with a field of 1T applied parallel to the crystallographic ab plane and (b) normalized electrical resistivity of the Lu(Fe$_{1-x}$Ru$_{x}$)$_{2}$Ge$_{2}$ series. Insets show data at low temperature. Transition temperatures are determined using the criteria described in the text.}%
\label{RuT}
\end{center}
\end{figure}

To a first order approximation, both Y and Sc substitutions cause only steric effects without changing the band filling. In an itinerant picture, another way to modify the sample without changing the band filling is to substitute Ru for Fe.  Figures \ref{RuT}a and \ref{RuT}b show the temperature dependent magnetization and resistivity data for the Lu(Fe$_{1-x}$Ru$_{x}$)$_{2}$Ge$_{2}$ series. It can be seen that by Ru substitution onto the Fe site the 8.2~K transition is suppressed. For {\itshape x} = 0.008, the lowest substitution level we were able to achieve, the anomaly in magnetization is suppressed to 5.2~K. The corresponding feature in resistivity is rather weak but can be seen clearly in the first derivative $dR/dT$ (not shown) giving a transition temperature of 4.6~K. For {\itshape x} = 0.014, there is an indication of drop in magnetization just as base temperature is approached; further, lower temperature measurements would be required to determine the precise transition temperature. No indication of a transition is observed in the resistivity data for this substitution level. For {\itshape x} = 0.035, neither magnetization nor resistivity data show any signs of a transition above 2~K.

\begin{figure}
\begin{center}
\resizebox*{14cm}{!}{\includegraphics{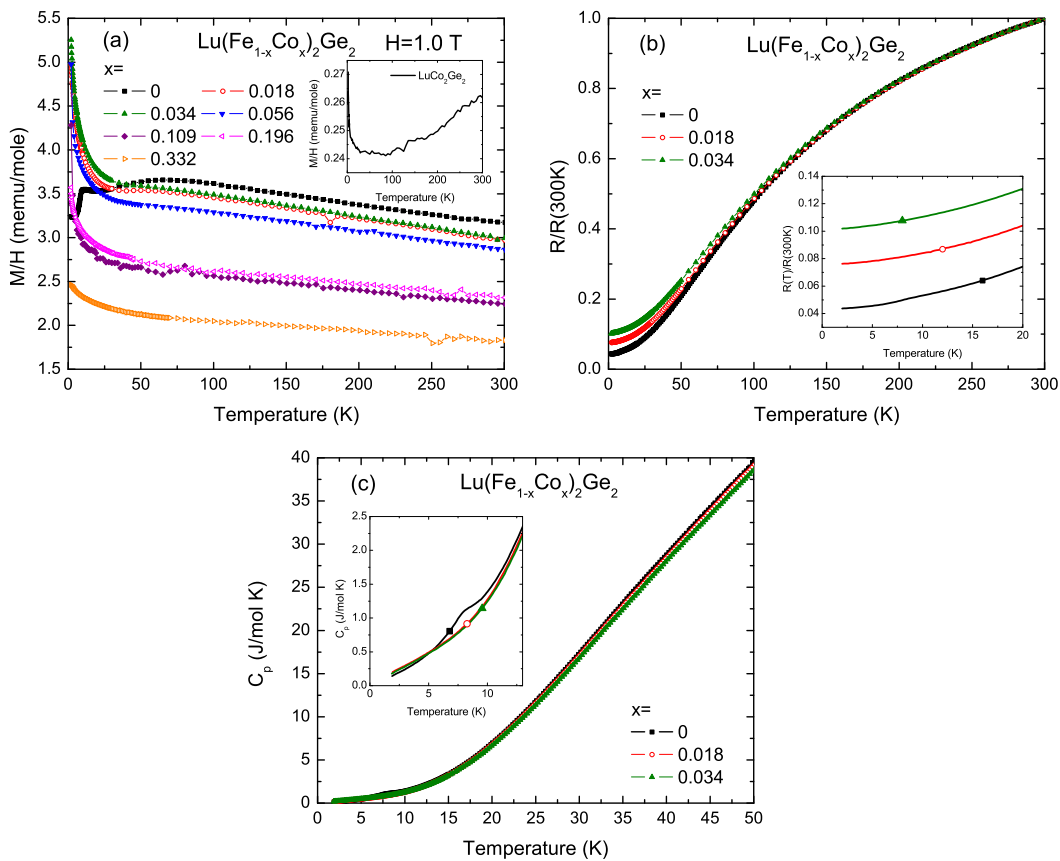}}%
\caption{ Temperature dependent (a) magnetization divided by applied field with a field of 1T applied parallel to the crystallographic ab plane, (b) normalized electrical resistivity and (c) specific heat of the Lu(Fe$_{1-x}$Co$_{x}$)$_{2}$Ge$_{2}$ series. Insets show data at low temperature. Transition temperatures are determined using the criteria described in the text.}%
\label{CoT}
\end{center}
\end{figure}

Whereas Y, Sc and Ru substitutions are expected to primarily cause only steric changes, Co substitution onto the Fe site, with one extra electron per atom, potentially affects the band filling. Figure \ref{CoT}a shows the temperature dependent magnetization data for the Lu(Fe$_{1-x}$Co$_{x}$)$_{2}$Ge$_{2}$ series. By Co substitution onto the Fe site, even with our lowest substitution level, {\itshape x} = 0.018, the anomaly is suppressed completely. With the magnetic transition suppressed, the magnetization data manifests the same type of upturn at low temperature that the Y and Ru substitutions data does.  As the Co substitution level is increased, the high temperature susceptibility decreases, consistent with the fact that LuCo$_{2}$Ge$_{2}$ has a susceptibility that is one order of magnitude smaller than that of LuFe$_{2}$Ge$_{2}$ (as shown in the inset to Figure \ref{CoT}a). The complete suppression of the 8.2 K feature by Co substitution is further confirmed by both resistivity and specific heat data which are shown in Figures \ref{CoT}b and \ref{CoT}c; neither the change of slope in resistivity nor the anomaly in the specific heat are detected in Co substituted compounds for any substitution levels. 

\begin{figure}
\begin{center}
\resizebox*{14cm}{!}{\includegraphics{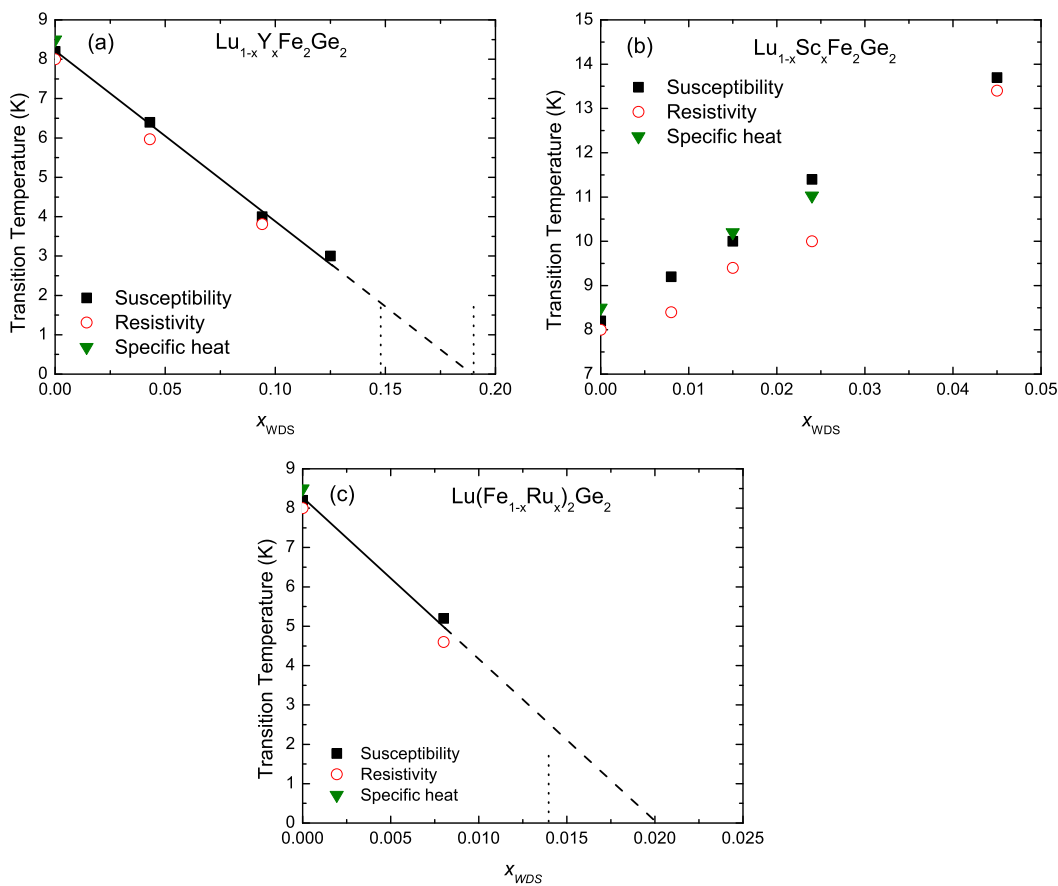}}
\caption{ T-x phase diagram for (a) the Lu$_{1-x}$Y$_{x}$Fe$_{2}$Ge$_{2}$ series, (b) the Lu$_{1-x}$Sc$_{x}$Fe$_{2}$Ge$_{2}$ series and (c) the Lu(Fe$_{1-x}$Ru$_{x}$)$_{2}$Ge$_{2}$ series. Squares are data from susceptibility data, circles are data from resistivity data, and triangles are data from specific heat data. Solid lines are the linear fit of the data. Dashed lines are the extrapolations of data to lower temperature. Vertical dotted lines represent the possible transitions below 2~K. }%
\label{Tx}
\end{center}
\end{figure}

Based on the magnetization, resistivity and specific heat data, the phase diagrams for Y, Sc and Ru substitutions are presented in Figure \ref{Tx}. The phase diagrams indicate a near linear suppression (enhancement) of the transition temperature for Y (Sc) substitution.  Ru substitution suppresses the transition at a higher rate than Y substitution and Co substitution suppresses the transition at least as rapidly as Ru substitution.

\begin{figure}
\begin{center}
\resizebox*{14cm}{!}{\includegraphics{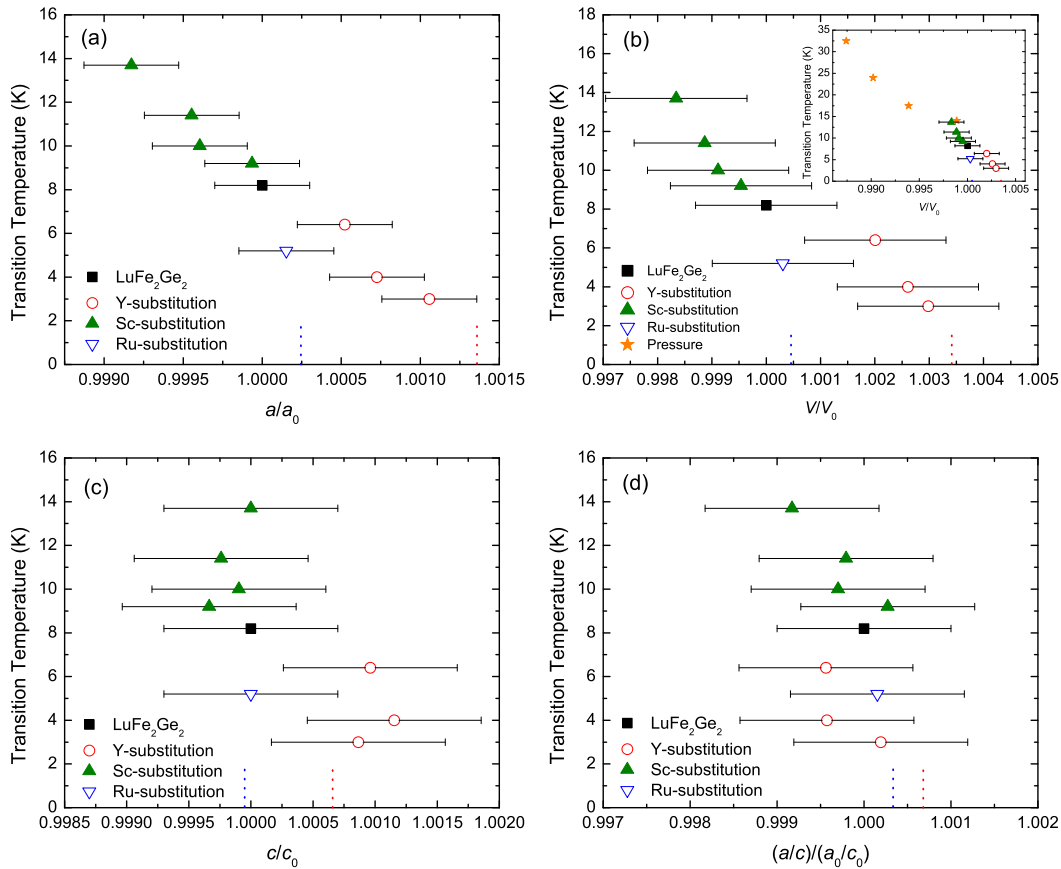}}%
\caption{Transition temperature versus normalized (a) lattice parameter {\itshape a}, (b) unit cell volume {\itshape V}, (c) lattice parameter {\itshape c} and (d) {\itshape a}/{\itshape c}. Stars in panel (b) are pressure data described in the text. Vertical dotted lines represent the possible transitions below 2~K.}%
\label{Tlattice}
\end{center}
\end{figure}

Given the apparent similarities between the effects of Sc substitution and applied pressure as well as the effects of Y and Ru substitutions it is worthwhile seeing if there is some underlying, unifying parameter that can be used to describe the effects of isoelectronic perturbations of the low temperature magnetic transition in LuFe$_{2}$Ge$_{2}$.  An examination of the plots in Figures \ref{lattice} and \ref{Tx} points to possible scaling of the transition temperature with either the unit cell volume or with the {\itshape a}-lattice parameter.  Figure \ref{Tlattice} presents the magnetic transition temperature as a function of {\itshape a}/{\itshape a}$_{0}$, {\itshape V}/{\itshape V}$_{0}$, {\itshape c}/{\itshape c}$_{0}$ and ({\itshape c}/{\itshape a})/({\itshape c}$_{0}$/{\itshape a}$_{0}$). Whereas changes in {\itshape a} and {\itshape V} may correlate with changes in the magnetic transition temperature, changes in {\itshape c} or {\itshape c}/{\itshape a} do not. Even though the changes in the {\itshape a}-lattice parameter (and volume) are rather small, some clear tendencies can still be extracted. It can be seen that for Y and Sc substitution, the transition temperatures can be well scaled with normalized {\itshape a}-lattice parameter, and to a lesser extent normalized volume (with Y substitution transition temperature values jump a little bit at first substitution level). It appears that Ru substitution with the higher rate at which it suppresses the transition temperature falls on the edge of the manifold for either normalized {\itshape a}-lattice parameter or normalized volume. The inset to Figure \ref{Tlattice}b includes the transition temperature data from LuFe$_{2}$Ge$_{2}$ under pressure.  In order to compare our data with those of the pressure study \cite{Fujiwara07}, the change of unit cell volume under pressure was estimated by using the bulk modulus of YbRh$_{2}$Si$_{2}$ \cite{Plessel03}, {\itshape B}$_{0}$ = 189~GPa, which is the closest compound that such data could be found for.  Considering the possibly differences between the bulk moduli of YbRh$_{2}$Si$_{2}$ and LuFe$_{2}$Ge$_{2}$, this is only a rough estimation. It appears that Y and Sc substitutions as well as the pressure data roughly follow the same scale of volume.  

\section{Conclusions}

The effects of Y and Sc substitutions for Lu, as well as Ru and Co substitutions for Fe, on the low temperature magnetic phase transition of LuFe$_{2}$Ge$_{2}$ have been studied in single crystals and the phase diagrams of Y, Sc and Ru substitution have been established. The results reveal that whereas Sc substitution, which serves as chemical pressure, enhances the transition, Y and Ru substitutions which serve as negative chemical pressure or strain suppress the transition to lower temperature. This is consistent with previous report of pressure study of the parent compound LuFe$_{2}$Ge$_{2}$ \cite{Fujiwara07}. In addition, for Y and Sc substitutions, there appear to be universal relations between transition temperature and both a-lattice parameter and volume so that transition temperature of these two substitutions can be scaled very well with both a-lattice parameter and volume. As this magnetic phase transition is suppressed no competing phase (such as superconductor) was revealed, at least for temperature above 2~K.

\section*{Acknowledgement}
The authors acknowledge A. Kreyssig and M. S. Torikachvili for valuable discussions and W. E. Straszheim for the elemental analysis of the crystals. K. Kandel is also acknowledged for helping shape one sample. Ames Laboratory is operated for the US Department of Energy by Iowa State University under Contract No. DE-AC02-07CH11358. This work was supported by the US Department of Energy, Office of Basic Energy Science, Division of Materials Sciences and Engineering. S.L.B. and P.C.C. were supported in part by the State of Iowa through the Iowa State University.


\label{lastpage}

\end{document}